\documentclass{elsart}

\usepackage{graphics}
\usepackage{graphicx}
\usepackage{epsfig}

\def\lambdabar {\mathchar'26\mkern-10mu\lambda}

\usepackage{amssymb}
\journal{Physics Letters A}

\begin{document}

\begin{frontmatter}


\title{A relativistic model of the isotropic three-dimensional singular oscillator}
\author{S.M. Nagiyev, E.I. Jafarov\corauthref{cor1}}
\ead{azhep@physics.ab.az}
\corauth[cor1]{Corresponding author}
\address{Institute of Physics, Azerbaijan National Academy of Sciences \\ Javid ave. 33, AZ1143, Baku, Azerbaijan}
\author{R.M. Imanov}
\address{Physics Department, Ganja State University \\ A. Camil str. 1, 374700, Ganja, Azerbaijan}
\author{L. Homorodean}
\address{University of Petrosani\\ University str. 20, 332006, Petrosani, Romania}





\begin{abstract}
Exactly solvable model of the quantum isotropic three-dimensional singular oscillator in the relativistic configurational $\vec r$-space is proposed. We have found the radial wavefunctions, which are expressed through the continuous dual Hahn polynomials and energy spectrum for the model under consideration. It is shown that they have the correct non-relativistic limits.
\end{abstract}

\end{frontmatter}

\section{Introduction}
\label{int}

The non-relativistic quantum harmonic oscillator \cite{landau} is extensively used in the various fields of the theoretical physics (see, for example, \cite{moshinsky}). The development of the quark models has led to the necessity of constructing the relativistic wavefunctions of compound particles and, in particular, the relativistic quantum harmonic oscillator models \cite{yukawa}-\cite{atakishiyev}.

Another useful solvable problem of the non-relativistic quantum mechanics is the singular harmonic oscillator \cite{frish}-\cite{castro} due to its application for explanation many fenomena, such as description of spin chains \cite{polychronakos}, quantum Hall effect \cite{frahm}, fractional statistics and anyons \cite{leinaas}.

Recently, we constructed a relativistic model of the quantum linear singular oscillator \cite{nagiyev}, which can be applied for studying relativistic physical systems as well as systems on a lattice. In the present paper we generalize this model to the three-dimensional case. Our three-dimensional model is formulated in the framework of the finite-difference relativistic quantum mechanics, which was developed in several papers and applied to the solution of a lot of problems in particle physics \cite{donkov,atakishiyev}, \cite{nagiyev}-\cite{frick}. 

In Section 2, we briefly discuss the finite-difference relativistic quantum mechanics. We propose a relativistic finite-difference model of the isotropic three-dimensional singular oscillator and find the explicit form of its wavefunctions and energy spectrum in Section 3. In Section 4, we investigate the non-relativistic limits of these wavefunctions and energy spectrum. Conclusions are given in Section 5.

\section{The finite-difference relativistic quantum mechanics}

The finite-difference relativistic quantum mechanics is closely analogous to the non-relativistic quantum mechanics, but its essential characteristic is that the relative motion wavefunction satisfies a finite-difference equation with a step equal to the Compton wavelength of the particle, $\lambdabar=\hbar/mc$. For example, in the case of a local quasipotential of interaction $V\left( \vec r \right) $ the equation for the wavefunction of two scalar particles with equal mass has the form

\begin{equation}
\label{4aa}
\left[ H_0+V\left( \vec r \right) \right] \psi (\vec r)=E\psi (\vec r) \; ,
\end{equation}
where the finite-difference operator $H_0$ is a relativistic free Hamiltonian

\begin{equation}
\label{5aa}
H_0=mc^2\left[ \cosh \left( i\lambdabar \partial _r\right) + \frac{i \lambdabar}r\sinh \left( i \lambdabar \partial _r\right) + \frac{  {\vec L}^2}{2\left( mcr \right) ^2}\exp \left( i\lambdabar \partial _r\right) \right]  \; ,
\end{equation}
and $\vec L^2$ is the square of the angular momentum operator and $\partial _r \equiv \frac \partial {\partial r}$. The technique of difference differentiation was developed and analogues of the important functions of the continuous analysis were obtained to fit the relativistic quantum mechanics, based on Eq. (\ref{4aa}) \cite{kadyshevsky,kadyshevsky2}.

The space of vectors $\vec r$ is called the relativistic configurational space or $\vec r$-space. The concept of $\vec r$-space has been introduced for the first time in the context of the quasipotential approach to the relativistic two-body problem \cite{kadyshevsky}.

The quasipotential equations for the relativistic scattering amplitude and the wave function $\psi \left( \vec p \right)$ in the momentum space have the form \cite{kadyshevsky4,kadyshevsky3}

\begin{eqnarray}
\label{5a}
A\left( \vec p,\vec q\right) =\frac m{4\pi }V\left( \vec p,\vec q;E_q\right) +\frac 1{\left( 2\pi \right) ^3}\int V\left( \vec p,\vec k;E_q\right) G_q\left(  k\right) A\left( \vec k,\vec q\right) d\Omega _k \; , \\
\label{6a}
\psi \left( \vec p\right) =\left( 2\pi \right) ^3\delta \left( \vec p\left( -\right) \vec q\right) +\frac 1{\left( 2\pi \right) ^3}G_q\left( p\right) \int V\left( \vec p,\vec k;E_q\right) \psi \left( \vec k\right) d\Omega _k \; ,
\end{eqnarray}
where

\begin{eqnarray}
\label{7a}
G_q\left( p\right) =\frac 1{E_q-E_p+i0},\quad \delta \left( \vec p\left( -\right) \vec q\right) \equiv \sqrt{1+\frac{\vec q^2}{m^2c^2}}\delta \left( \vec p-\vec q\right) , \\
d\Omega _k=\frac{d\vec k}{\sqrt{1+\frac{\vec k^2}{m^2c^2}}},\quad E_q=\sqrt{\vec q^2c^2+m^2c^4}, \nonumber
\end{eqnarray}
and $V\left(\vec p, \vec k; E_q \right)$ is the quasipotential.

The integration in (\ref{5a}) and (\ref{6a}) is carried out over the mass shell of the particle with mass $m$, i.e. over the upper sheet of the hyperboloid ${p_0}^2-{\vec p}^2=m^2c^2$, which from the geometrical point of view realizes the three-dimensional Lobachevsky space. The group of motions of this space is the Lorentz group $SO(3,1)$.

Equations (\ref{5a}) and (\ref{6a}) have the absolute character with respect to the geometry of the momentum space, i.e., formally they don't differ from the non-relativistic Lippmann-Schwinger and Schr\"odinger equations. We can derive Eqs. (\ref{5a}) and (\ref{6a}) substituting the relativistic (non-Euclidean) expressions for the energy, volume element, and $\delta$-function by their non-relativistic (Euclidean) analogues:

\begin{eqnarray}
\label{8a}
E_q=\frac{q^2}{2m} \rightarrow E_q = \sqrt{{\vec q}^2 c^2 +m^2 c^4} \; , \nonumber \\
d \vec k \rightarrow d \Omega _k = \frac{d\vec k}{\sqrt{1+\frac{{\vec k}^2}{m^2c^2}}}\;, \\
\delta\left(\vec p - \vec q \right) \rightarrow \delta\left(\vec p (-) \vec q \right). \nonumber
\end{eqnarray}

As a consequence of this geometrical treatment, the application of the Fourier transformation to the Lorentz group becomes natural instead of the usual one. In this case the relativistic configurational $\vec r$-space conseption arises.

Transition to relativistic configurational $\vec r$-representation

\begin{equation}
\label{9a}
\psi \left( \vec r\right) =\frac 1{\left( 2\pi \hbar \right) ^{\frac 32}}\int \xi \left( \vec p,\vec r\right) \psi \left( \vec p\right) d\Omega _p
\end{equation}
is performed by the use of expansion on the matrix elements of the principal series of the unitary irreducible representations of the Lorentz group:

\begin{eqnarray}
\label{10a}
\xi \left( \vec p,\vec r\right) =\left( \frac{p_0-\vec p\vec n}{mc}\right) ^{-1-ir/\lambdabar } \; ,  \\
\vec r=r\vec n,\quad 0\leq r<\infty ,\nonumber \\ \vec n=\left( \sin \theta \cos \varphi ,\sin \theta \sin \varphi ,\cos \theta \right) ,\; p_0=\sqrt{\vec p^2+m^2c^2}. \nonumber
\end{eqnarray}

The quantity $r$ is relativistic invariant and is connected with the eigenvalues of the Casimir operator $\hat C = \vec N ^2- \vec L ^2$ in the following way:

\begin{equation}
\label{11a}
C=\lambdabar  ^2 + r ^2 \; ,
\end{equation}
where $\vec L$ and $\vec N$ are the rotation and boost generators.

It is easy to verify that the function (the relativistic 'plane wave') (\ref{10a}) obeys the finite-difference Schr\"odinger equation

\begin{equation}
\label{12a}
\left(H_0-E_p \right) \xi \left(\vec p, \vec r \right) = 0 \; .
\end{equation}

The relativistic plane waves form a complete and orthogonal system of functions in the momentum Lobachevsky space.

If we perform the relativistic Fourier transformation (\ref{9a}) in Eq. (\ref{6a}), we arrive at the finite-difference Schr\"odinger equation (\ref{4aa}) with the local (in general case, non-local) potential in the relativistic $\vec r$-space.

In the relativistic $\vec r$-space the Euclidean geometry is realized and, in particular, there exists a momentum operator in the relativistic configurational $\vec r$ representation \cite{kadyshevsky}

\begin{equation}
\label{13a}
\hat {\vec p} = - mc \vec n \left( e ^{i \lambdabar \partial _r} - \frac{H_0}{mc^2} \right) - \vec m \frac{\hbar}{r} e ^{i \lambdabar \partial _r} \; ,
\end{equation}
where a three-dimensional vector $\vec m$ has the following components \cite{nagiyev2}:

\begin{eqnarray}
m_1=i\left( \cos \varphi \cos \theta \frac \partial {\partial \theta }-\frac{\sin \varphi }{\sin \theta }\frac \partial {\partial \varphi }\right) \; , \nonumber \\
m_2=i\left( \sin \varphi \cos \theta \frac \partial {\partial \theta }-\frac{\cos \varphi }{\sin \theta }\frac \partial {\partial \varphi }\right) \; , \nonumber \\
m_3=-i\sin \theta \frac \partial {\partial \theta } \; .
\end{eqnarray}

The components of (\ref{13a}) and free Hamiltonian obey the following commutation relations:

$$
\left[ \hat p_i,\hat p_j\right] =\left[ \hat p_i,H_0\right] =0,\quad i,j=1,2,3 \; .
$$

The relativistic plane wave is the eigenfunction of the operator $\hat {\vec p}$:

\begin{equation}
\label{14a}
\hat {\vec p} \xi \left(\vec p, \vec r \right) = \vec p \xi \left(\vec p, \vec r \right) \; .
\end{equation}

This means that (\ref{10a}) describes the free relativistic motion with definite energy and momentum.

In the non-relativistic limit we come to the usual three-dimensional configurational space and relativistic plane wave (\ref{10a}) goes over into the Euclidean plane wave, i.e.

\begin{equation}
\label{8}
\lim _{c \rightarrow \infty} \xi \left( \vec p, \vec r \right) = e ^{i \vec p \vec r / \hbar} \; .
\end{equation}

Note that all the important exactly solvable cases of non-relativistic quantum mechanics (potential well, Coulomb potential, harmonic oscillator etc.) are also exactly solvable for the case of Eq. (\ref{4aa}).

\section{The relativistic model of the isotropic three-dimensional singular oscillator}

We consider a model of the relativistic three-dimensional singular oscillator, which corresponds to the following interaction quasipotential:

\begin{equation}
\label{eq5}
V\left( {\vec r} \right) = \left\{ {\frac{1}{2}m\omega ^2  \left(r+i\lambdabar \right)^2  + \frac{g}{{r^2 }}} \right\}e^{i\lambdabar \partial _r }, 
\end{equation}
where $g$ is a real quantity.

Let us note that in contrast to the case of the Coulomb potential \cite{nagiyev,freeman,nagiyev2}, which can be calculated as an input of the one-photon exchange, the relativistic generalization of the oscillator or singular oscillator potential is not uniquely defined. Therefore, for construction of the quasipotential (\ref{eq5}) we proceed from the following requirements for the quasipotential: a) exact solubility; b) the correct non-relativistic limit; c) existence of the dynamical symmetry.

In the limiting case when $c \rightarrow \infty$, $V \left( \vec r \right)$ goes into the non-relativistic three-dimensional singular harmonic oscillator potential \cite{landau}:

\[
V(r) \rightarrow \frac 12 m \omega ^2 r ^2 + \frac g{r^2}.
\]

Due to spherical symmetry of (\ref{eq5}) the angular dependences of the wavefunction $\psi \left( \vec r \right)$ (\ref{4aa}) are separated in the standard manner

\begin{equation}
\label{eq6}
\psi \left( {\vec r} \right) = \frac{1}{r}R_l \left( r \right)Y_{lm} \left( {\theta ,\varphi } \right),
\end{equation}
where $l=0,1,2, \dots$ being the orbital quantum number.

Thus the three-dimensional problem is reduced to finding the eigenvalues and eigenfunctions of the radial part of a Hamiltonian

\begin{equation}
\label{eq7}
H_l R_l (r) = E_l R_l (r)
\end{equation}
with the boundary conditions for the radial wavefunction $R_l(0)=R_l (\infty)=0$, where

\begin{equation}
\label{eq8}
H_l  = mc^2 \left[ {\cosh \left( {i\lambdabar \partial _r } \right) + \frac {\lambdabar ^2 l(l + 1)}{2r\left( {r + i\lambdabar } \right)}e^{i\lambdabar \partial _r } } \right] + \left[ {\frac{1}{2}m\omega ^2 r\left( {r + i\lambdabar } \right) + \frac{g}{{r\left( {r + i\lambdabar } \right)}}} \right]e^{i\lambdabar \partial _r }.
\end{equation}

In terms of dimensionless quantities $\rho=r/\lambdabar$, $\omega_0=\hbar \omega/mc^2$ and $g_0=mg/\hbar^2$ we can write equation (\ref{eq7}) in the form

\begin{equation}
\label{eq9}
\left[ {\cosh \left( {i\partial _\rho  } \right) + \frac{1}{2}\omega _0 ^2 \rho ^{(2)} e^{i\partial _\rho  }  + \frac{{2g_0  + l(l + 1)}}{{2\rho ^{(2)} }}e^{i\partial _\rho  } } \right]R_l \left( \rho  \right) = \frac{{E_l }}{{mc^2 }}R_l \left( \rho  \right)
\end{equation}
where $\rho^{(2)}$ is the "generalized degree" \cite{kadyshevsky2}

\begin{equation}
\label{eq10}
\rho ^{\left( \delta  \right)}  = i^\delta  \frac{{\Gamma \left( {\delta  - i\rho } \right)}}{{\Gamma \left( { - i\rho } \right)}}.
\end{equation}

Having the seperated factors $\left( - \rho \right) ^ {\left( \alpha_l \right)}$ and $M_{\nu_l}\left(\rho \right) = \omega _0 ^{i \rho} \Gamma \left( \nu _l + i \rho \right)$ with

\begin{eqnarray}
\label{eq11}
\alpha _l  = \frac{1}{2} + \frac{1}{2}\sqrt {1 + \frac{2}{{\omega _0 ^2 }}\left( {1 - \sqrt {1 - 8g_0 \omega _0 ^2  - 4\omega _0 ^2 l\left( {l + 1} \right)} } \right)}, \\
\label{eq12}
\nu _l  = \frac{1}{2} + \frac{1}{2}\sqrt {1 + \frac{2}{{\omega _0 ^2 }}\left( {1 + \sqrt {1 - 8g_0 \omega _0 ^2  - 4\omega _0 ^2 l\left( {l + 1} \right)} } \right)},
\end{eqnarray}
which determine the asymptotic behaviour of $R_l \left( \rho \right)$ at the points $\rho=0$ and $\rho=\infty$ respectively, we look for the solution of equation (\ref{eq9}) in the form

\begin{equation}
\label{eq13}
R\left( \rho  \right) = C_l \left( { - \rho } \right)^{\left( \alpha_l  \right)} M_{\nu_l} \left( \rho \right) \Omega \left( \rho  \right).
\end{equation}

The function $\Omega \left( \rho \right)$ then satisfies the following difference equation

\begin{equation}
\label{eq14}
\left[ {\left( {\alpha _l  + i\rho } \right)\left( {\nu _l  + i\rho } \right)e^{ - i\partial _\rho  }  - \left( {\alpha _l  - i\rho } \right)\left( {\nu _l  - i\rho } \right)e^{i\partial _\rho  } } \right]\Omega \left( \rho  \right) = 2i\frac{{E_l }}{{\hbar \omega }}\Omega \left( \rho  \right).
\end{equation}

This equation was studied in \cite{atakishiyev,nagiyev}, where has been shown that the insertion of expansion of the function $\Omega \left( \rho  \right)$ as a power series into (\ref{eq14}) leads to recurrence relations for the coefficients of this series, from which it follows that it terminates by the term $c_{2n} \rho ^{2n}$ if

\begin{equation}
\label{eq15}
E_l \equiv E_{nl} = \hbar \omega \left( 2n +\alpha _l + \nu_l \right) ,
\end{equation}
where $n=0,1,2, \dots $ is the radial quantum number. This gives the quantization rule for the energy levels of the three-dimensional singular oscillator under consideration and we arrive at the following final form for the radial wavefuncion

\begin{equation}
\label{eq16}
R_{nl} \left( \rho  \right) = C_{nl} \left( { - \rho } \right)^{\left( \alpha_l  \right)} M_{\nu_l} \left( \rho \right) S_n \left(\rho ^2; \alpha _l, \nu _l, \frac 12 \right),
\end{equation}
where $S_n \left(\rho ^2; \alpha _l, \nu _l, \frac 12 \right)$ are the continuous dual Hahn polynomials \cite{koekoek}.

If we normalize the functions (\ref{eq16}) as follows

\begin{equation}
\label{eq17}
\int \limits _0 ^\infty R_{nl} \left( \rho  \right) R_{ml}^* \left( \rho  \right) \d \rho = \delta _{nm},
\end{equation}
then
\[
C_{nl}  = \sqrt {\frac{2}{{n!\Gamma \left( {n + \alpha _l  + \nu _l } \right)\Gamma \left( {n + \alpha _l  + 1/2} \right)\Gamma \left( {n + \nu _l  + 1/2} \right)}}}. 
\]

\section{The non-relativistic limit case}

In the non-relativistic limit, when the parameter $\mu=\frac {mc^2}{\hbar \omega}\rightarrow \infty$ we have \cite{nagiyev}

\[
\begin{array}{l}
 \mathop {\lim }\limits_{\mu  \to \infty } \alpha _l  = \frac{1}{2} + \frac{1}{2}\sqrt {\left( {2l + 1} \right)^2  + \frac{{8mg}}{{\hbar ^2 }}}  \equiv 2s + 1, \\ 
 \mathop {\lim }\limits_{\mu  \to \infty } \left( {\nu _l  - \mu } \right) = \frac{1}{2}, \\ 
 \mathop {\lim }\limits_{\mu  \to \infty } \left( { - \rho } \right)^{\left( {\alpha _l } \right)}  = e^{\left( {s + \frac{1}{2}} \right)\ln \mu } \left( { - \xi } \right)^{2s + 1} , \\ 
 \mathop {\lim }\limits_{\mu  \to \infty } M_{\nu _l } \left( \rho  \right) = \sqrt {2\pi } e^{\mu \ln \mu  - \mu  - \frac{{\xi ^2 }}{2}} , \\ 
 \mathop {\lim }\limits_{\mu  \to \infty } C_{nl}  = \frac{1}{{\sqrt {\pi n!\Gamma \left( {n + 2s + 3/2} \right)} }}e^{\mu  - \left( {\mu  + n + s + \frac{3}{4}} \right)\ln \mu } , \\ 
 \mathop {\lim }\limits_{\mu  \to \infty } \frac{1}{{n!\mu ^n }}S_n \left( {\rho ^2 ;\alpha _l ,\nu _l ,1/2} \right) = L_n^{2s + 1/2} \left( {\xi ^2 } \right), \\ 
 \end{array}
\]
where $\xi = \sqrt{\frac{m\omega}\hbar }r$. Therefore in the limit $c \rightarrow \infty$  the radial wavefunctions $R_{nl} \left( \rho \right)$ (\ref{eq16}) coincide with the radial wavefunctions of the non-relativistic three-dimensional singular oscillator \cite{landau}.

\begin{figure}[h!]
\begin{minipage}[b]{0.5\linewidth} 
\epsfig{file=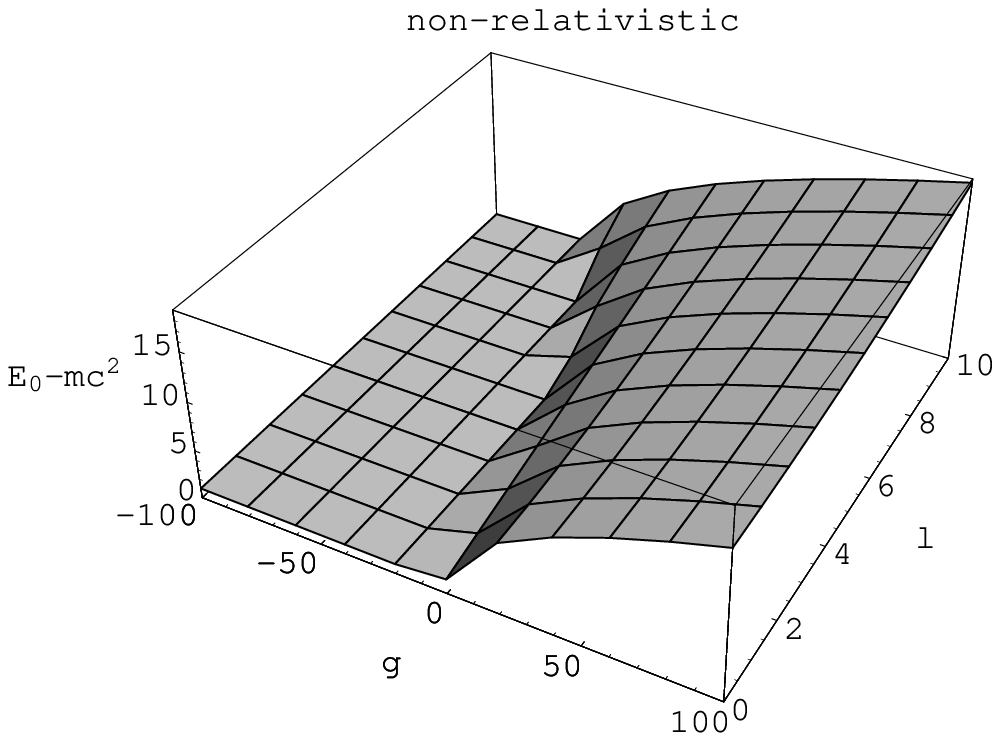,height=5.36cm,width=6.36cm}
\end{minipage}
\hspace{0.5cm} 
\begin{minipage}[b]{0.5\linewidth}
\epsfig{file=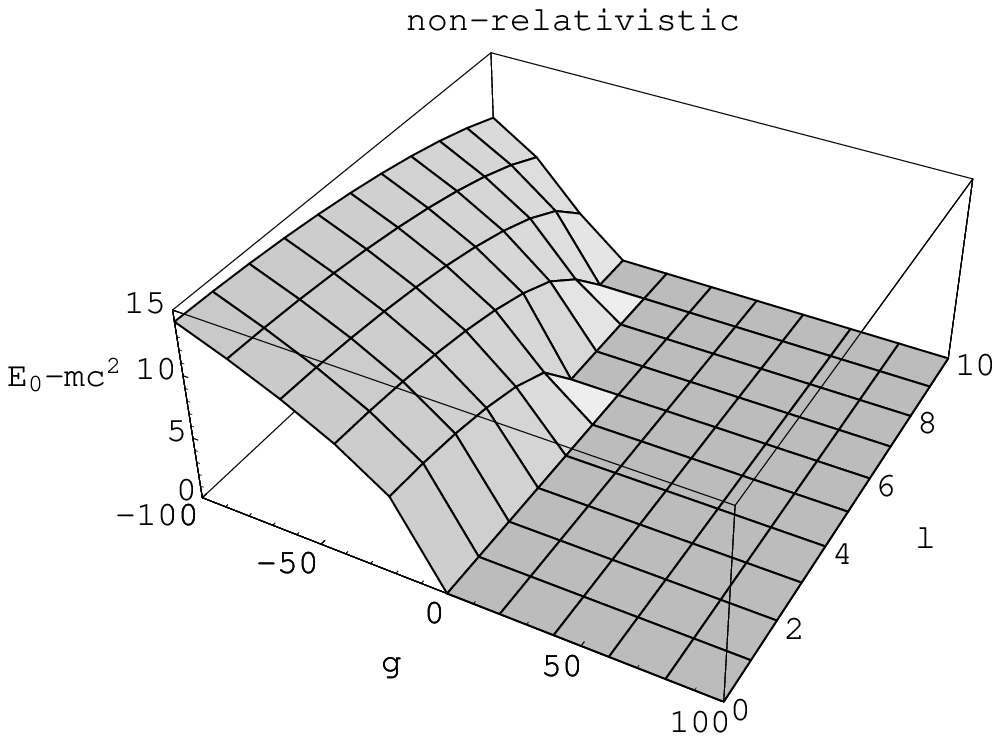,height=5.36cm,width=6.36cm}
\end{minipage}
\vspace{1.5cm}
\begin{minipage}[b]{0.5\linewidth} 
\epsfig{file=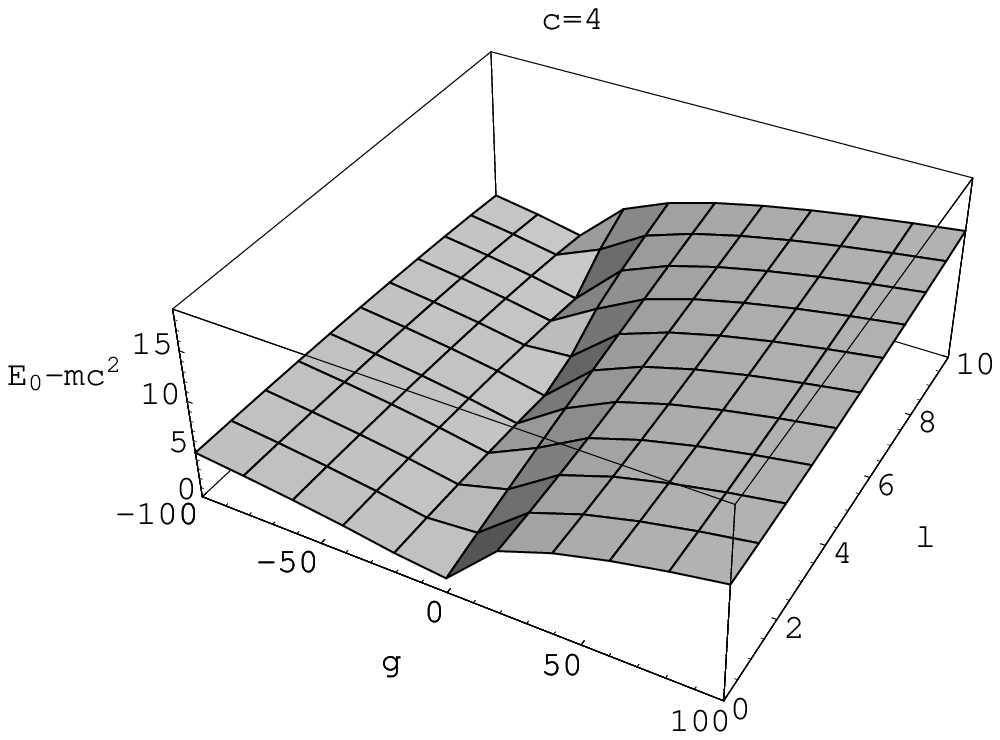,height=5.36cm,width=6.36cm}
\end{minipage}
\hspace{0.5cm} 
\begin{minipage}[b]{0.5\linewidth}
\epsfig{file=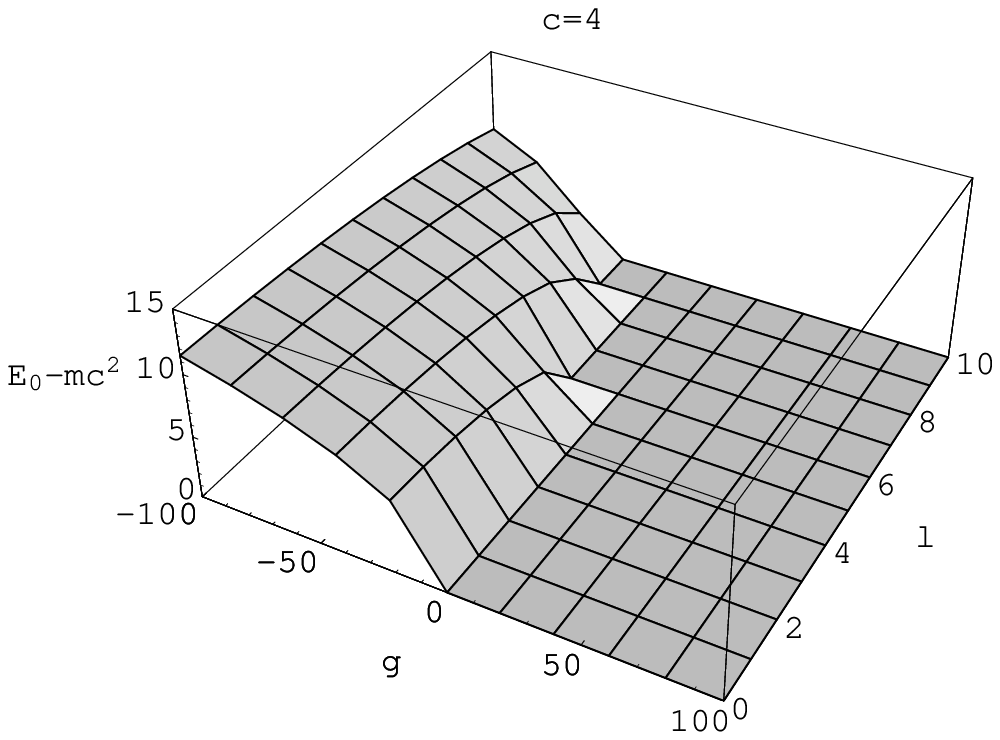,height=5.36cm,width=6.36cm}
\end{minipage}
\begin{minipage}[b]{0.5\linewidth} 
\epsfig{file=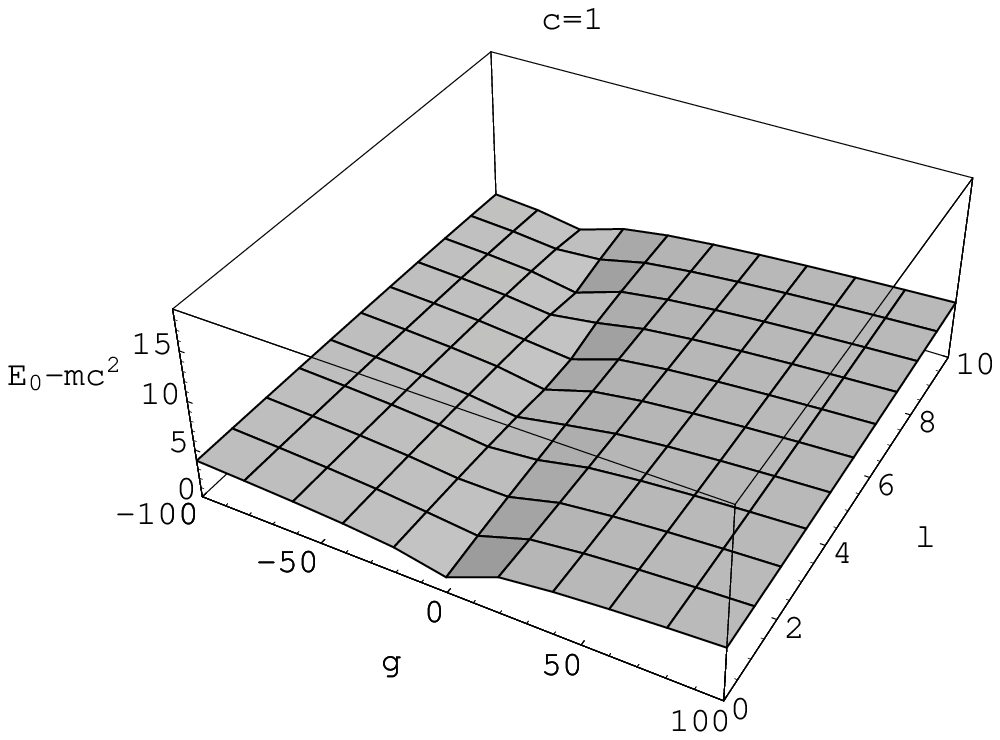,height=5.36cm,width=6.36cm}
\end{minipage}
\hspace{0.5cm} 
\begin{minipage}[b]{0.5\linewidth}
\epsfig{file=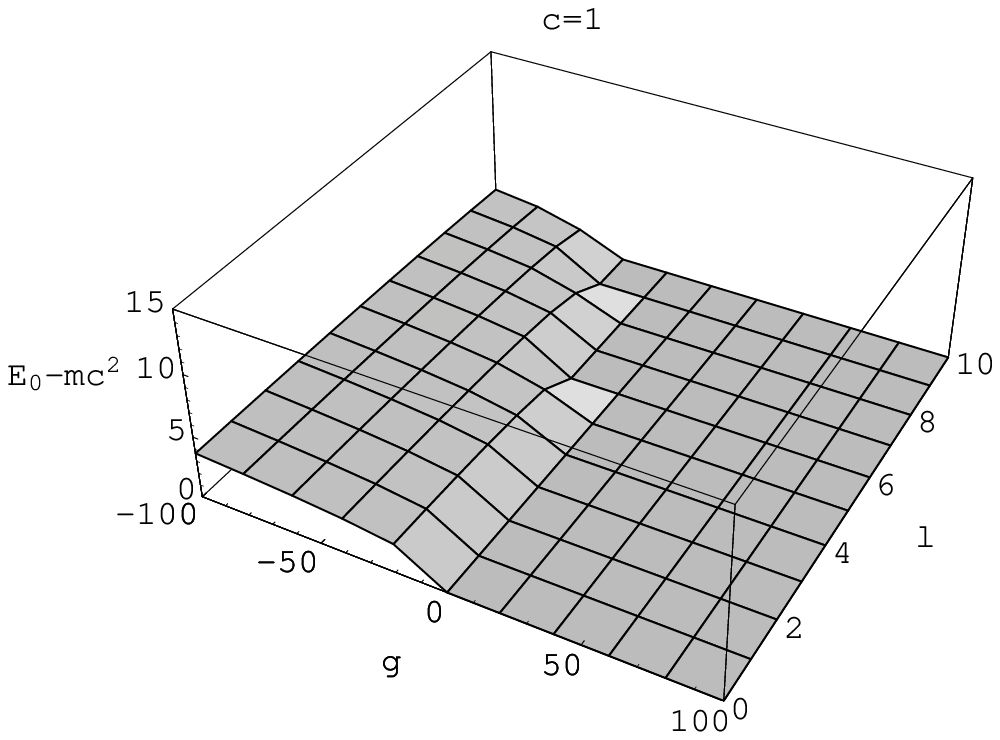,height=5.36cm,width=6.36cm}
\end{minipage}
\caption{\label{fig1} The behaviour of the ground state energy-level of the 3D relativistic isotropic singular oscillator (\ref{eq15}) for values of speed of light $c= \infty$ (the non-relativistic case), $4$ and $1$ ($m=\omega=\hbar=1$). Real parts are shown in the left plots, imaginary parts in the right plots.}
\end{figure}
The energy spectrum (\ref{eq15}) also has a correct non-relativistic limit, i.e.

\[
\begin{array}{l}
\mathop {\lim }\limits_{c \to \infty } \left( {E_n  - mc^2 } \right) = E_n ^{{\rm non - rel}}  = \hbar \omega \left( {2n + 2s + 3/2} \right) \\  = \hbar \omega \left( {2n + 1 + \frac{1}{2}\sqrt {\left( {2l + 1} \right)^2  + \frac{{8mg}}{{\hbar ^2 }}} } \right). \\ 
\end{array}
\]

In fig. 1 we show the behaviour of the ground state energy level (\ref{eq15}) depending on the $g$ and $l$ for various values of the speed of light $c$. From these plots we see that decreasing $c$ and increasing $l$ changes the appearance of the 'collapse' point $g< - \frac {\hbar^2}{8m} - \frac {\hbar ^2 \omega_0 ^2}{32m} - \frac {\hbar^2}{2m} l(l+1)$ \cite{landau,dodonov,nagiyev}.

\section{Conclusion}

In this paper in the framework of the finite-difference relativistic quantum mechanics we constructed a relativistic model of the isotropic three-dimensional singular oscillator. In complete analogy with the non-relativistic problem, the relativistic problem under consideration is also exactly solvable. We determined energy spectrum and wavefunctions of the problem and showed that they have the correct non-relativistic limits.

We hope that the relativistic model of the isotropic three-dimensional singular oscillator proposed in this paper will be applied in future in quantum physics as well as theory of elementary particles.




\begin{thebibliography}{00}




\bibitem{landau}
L.D. Landau and  E.M. Lifshitz, {\it Quantum Mechanics: Non-Relativistic Theory} (Oxford: Butterworth-Heinemann), 1997.

\bibitem{moshinsky}
M. Moshinsky, {\it The Harmonic Oscillator in Modern Physics: from Atoms ro Quarks} (New-York: Gordon and Breach), 1969.

\bibitem{yukawa}
H. Yukawa, Phys. Rev. \textbf{91} (1953) 416-417.

\bibitem{feynman}
R.P. Feynman, M. Kislinger and F. Ravndal, Phys. Rev. D \textbf{3} (1971) 2706-2732.

\bibitem{kim}
Y.S. Kim and M.E. Noz, Amer. J. Phys. \textbf{46} (1978) 480-483.

\bibitem{donkov}
A.D. Donkov, V.G. Kadyshevsky, M.D. Mateev and R.M. Mir-Kasimov, Teor. Mat. Fiz. \textbf{8} (1971) 61-68.

\bibitem{atakishiyev}
N.M. Atakishiyev, R.M. Mir-Kasimov and S.M. Nagiyev, Theor. Math. Phys. \textbf{44} (1980) 592-603; N.M. Atakishiyev, R.M. Mir-Kasimov and S.M. Nagiyev, Ann. Phys., Lpz \textbf{42} (1985) 25-30.

\bibitem{frish}
J. Frish, V. Mandrosov, Ya.A. Smorodinsky, M. Uhlir and P. Winternitz, Phys. Lett. \textbf{16} (1965) 354-356.

\bibitem{calogero}
F. Calogero, J. Math. Phys. \textbf{10} (1969) 2191-2196.

\bibitem{dodonov}
V.V. Dodonov, I.A. Malkin and V.I. Man'ko, Physica \textbf{72} (1974) 597-615.

\bibitem{dodonov2}
V.V. Dodonov, V.I. Man'ko and L. Rosa, Phys. Rev. A \textbf{57} (1998) 2851-2858.

\bibitem{yuce}
C. Yuce, Ann. Phys.  \textbf{308} (2003) 599-604; Phys. Lett. A \textbf{321} (2004) 291-294.

\bibitem{castro}
A.S. de Castro, Ann. Phys. \textbf{311} (2004) 170-181.

\bibitem{polychronakos}
A.P. Polychronakos,  Phys. Rev. Lett. \textbf{69} (1992) 703-705.

\bibitem{frahm}
H. Frahm, J. Phys. A: Math. Gen. \textbf{26} (1993) L473-L479.

\bibitem{leinaas}
J.M. Leinaas and J. Myrheim, Phys. Rev. B \textbf{37} (1988) 9286-9291.

\bibitem{nagiyev}
S.M. Nagiyev, E.I. Jafarov and R.M. Imanov, J. Phys. A: Math. Gen. \textbf{36} (2003) 7813-7824.

\bibitem{kadyshevsky}
V.G. Kadyshevsky, M.R. Mir-Kasimov and N.B. Skachkov, Nuovo Chim. \textbf{A55} (1968) 233-257.

\bibitem{kadyshevsky2}
V.G. Kadyshevsky, M.R. Mir-Kasimov and M. Freeman, Yad. Fiz. \textbf{9} (1969) 646-652.

\bibitem{freeman}
M. Freeman, M.D. Mateev and R.M. Mir-Kasimov, Nucl. Phys. B \textbf{12} (1969) 197-215.

\bibitem{kadyshevsky4}
V.G. Kadyshevsky, Nucl. Phys. \textbf{6} (1968) 125-148.

\bibitem{kadyshevsky3}
V.G. Kadyshevsky, R.M. Mir-Kasimov and N.B. Skachkov, Phys. Elem. Part. At. Nucl. \textbf{2} (1971) 635-690.

\bibitem{jhung}
K.S. Jhung, K.H. Chung and R.S. Willey, Phys. Rev. D \textbf{12} (1975) 1999-2001.

\bibitem{kagramanov}
E.D. Kagramanov, R.M. Mir-Kasimov and S.M. Nagiyev, J. Math. Phys. \textbf{31} (1990) 1733-1738.

\bibitem{nagiyev2}
S.M. Nagiyev,  J. Phys. A: Math. Gen. \textbf{21} (1988) 2559-2564.

\bibitem{kagramanov2}
E.D. Kagramanov, R.M. Mir-Kasimov and S.M. Nagiyev, Phys. Lett. A \textbf{140} (1989) 1-4.

\bibitem{milton}
K.A. Milton and I.L. Solovtsov, Mod. Phys. Lett. A \textbf{16} (2001) 2213-2219.

\bibitem{frick}
R.A. Frick, Eur. Phys. J. C \textbf{28} (2003) 431-435. 

\bibitem{koekoek}
R. Koekoek and R.F. Swarttouw, {\it Delft University of Technology}, {Report no. \bf 98-17} (1998).

\end{thebibliography}
\end{document}